\documentclass{article}
\usepackage{amssymb}

\usepackage{graphicx}
\usepackage{amsmath}

\input{tcilatex}

\begin{document}

\title{Antiferromagnetism of almost localized fermions: \\
Evolution from Slater-type to Mott-Hubbard gap}
\author{P. Korbel,$\;$ W. W\'{o}jcik \\
Institute of Physics, Technical University, \\
ulica Podchor\c{a}\.{z}ych 1, 30-084 Krak\'{o}w, Poland \and A. Klejnberg, $%
\;$ J. Spal{}ek \\
Marian Smoluchowski Institute of Physics, \\
Jagiellonian University, ulica Reymonta 4, \\
30-059 Krak\'{o}w, Poland \and M. Acquarone \\
C.N.R.-I.M.E.M., Unita I.N.F.M. di Parma,\\
Dipartimento di Fisica, Universita' di Parma, \\
Parco Area delle Scienze 7A, 43100 Parma, Italy \and and $\;$ M. Lavagna \\
Commissariat \'{a} l'Energie Atomique, \\
DRFMC /SPSMS, rue des Martyrs, \\
38054 Grenoble Cedex 9, France}
\maketitle

\begin{abstract}
We supplement (and critically overview) the existing extensive analysis of
antiferromagnetic solution for the Hubbard model with a detailed discussion
of two specific features, namely (i) the evolution of the magnetic (Slater)
gap (here renormalized by the electronic correlations) into the Mott-Hubbard
or atomic gap, and (ii) a rather weak renormalization of the effective mass
by the correlations in the half-filled-band case, which contrasts with that
for the paramagnetic case. The mass remains strongly enhanced in the
non-half-filled-band case. We also stress the difference between magnetic
and non-magnetic contributions to the gap. These results are discussed
within the slave boson approach in the saddle-point approximation, in which
there appears a \textit{non-linear }staggered molecular field due to the
electronic correlations that leads to the appearance of the \textit{magnetic
gap}. They reproduce correctly the ground-state energy in the limit of
strong correlations. A brief comparison with the solution in the limit of
infinite dimensions and the corresponding situation in the
doubly-degenerate-band case with one electron per atom is also made.

PACS \ \ Nos. 71.10.Fd, 75.10.Lp, 75.50.Ee
\end{abstract}

\section{Introduction}

Antiferromagnetism (AF) appears in the Hubbard and related models for an
arbitrary interaction strength $U$ provided we are close to the
half-filled-band situation ($n\rightarrow1$ in the orbitally nondegenerate
case) \cite{1}. This is easy to understand qualitatively, since the
intraatomic Hubbard interaction $U\sum_{i}{}\langle n_{i\uparrow
}n_{i\downarrow}\rangle$ is diminished by keeping apart the electrons with
the opposite spins. In the strong-correlation limit \cite{2} this
interaction leads to an antiferromagnetic kinetic exchange for an arbitrary
band filling $n$. At the same time, the band energy is not increased because
the concomitant nesting condition for the quasiparticle states achievable
for bipartite lattices only leads to the energy decrease of the occupied
states, even when going beyond the Hartree-Fock picture. In effect, the
regime of the band filling $n,$ for which the AF state is stable at given $U$
has been determined for variety of theoretical approaches \cite{1} - \cite{4}%
. The reliability of the results for the half-filled case is not in question
as long as they reduce to those in the Hartree-Fock and to the mean-field
(Heisenberg) approximations in the weak- and strong-correlation limits,
respectively, as we will discuss in the following. The theoretical results
are in accord with the fact that \textit{all} known Mott insulators with the
half-filled band configuration are also antiferromagnetic insulators. The
main purpose of this paper is to overview the situation by discussing the
crossover behavior from the Hartree-Fock to the mean-field approximation for
the Heisenberg antiferromagnet. Similar, though not equivalent results can
be reached within the dynamic mean-field theory (DMFT, see below). In the
orbitally degenerate situation (with the degeneracy $d>1$) the same type of
magnetic(Slater) gap is generated by an alternant orbital ordering \cite{3},
as is also discussed at the end of the paper.

Explicitly, we concentrate our attention to two specific features of
quasiparticle states not elaborated in detail so far, namely, (i) an
evolution of the magnetic gap (renorma\-li\-zed by the electronic
correlations) into the Mott-Hubbard gap, and (ii) a rather weak
renormalization of the effective mass for the half-filled-band case, which
is in contrast with that calculated in the paramagnetic (PARA) case \cite{4}%
. A rather strong mass enhancement is retained in the $n\neq 1$ case, in
direct analogy to the paramagnetic case \cite{5}. These results are obtained
within the slave-boson approach in the saddle-point approximation, which we
compare with the corresponding analysis in the infinite-dimension limit \cite
{1}. In particular, we introduce the concept of a nonlinear staggered
molecular field, which shows up as the effective
(nonlinear-in-magnetization) magnetic gap, evolving at temperature $T=0$
continuously with increasing $U$ from the Slater (Hartree-Fock) gap $(\sim
U) $ into the Hubbard gap. In connection with this evolution, we single out
the magnetic and Coulomb parts of the localization energy. These particular
features resolve explicitly the old question about the difference between
the Slater and Mott-Hubbard insulators in the sense that only the
Mott-Hubbard gap survives when antiferromagnetism disappears (i.e. above the
N\'{e}el temperature). Most of the results are contained in the mathematical
formulation established earlier \cite{1} - \cite{6}. Here, we only discuss
those points in an explicit manner. We believe that these points are
relevant to the general physics of correlated systems. This reason is also
behind publishing, perhaps belatedly, such a simple paper. Additionally,
since the effective mass is only weakly renormalized for the $n=1$ case and
it never reaches zero for $n\neq 1$, the role of the quantum fluctuations is
relatively not as crucial as for the discussion of the paramagnetic state.

The structure of the paper is as follows. In Section II we provide the
analytic details of the AF solution, which are followed by the numerical
analysis (Section III) and the discussion in physical terms in Section IV,
where we also compare the results with those for a doubly degenerate-band
case for $n=1$.

\section{Antiferromagnetic solution}

\subsection{{\protect\large \textbf{Saddle-point approximation}}}

To stress the role of the molecular field coming from the electronic
correlations, we start from the extended Hubbard model, with intersite
exchange interactions included, i.e. write 
\begin{equation}
H=\underset{\QATOP{i\in A}{j\in B}}{\sum}t_{ij}a_{i\sigma}^{\dagger}a_{j%
\sigma}+U\underset{i\in A}{\sum }n_{i_{\uparrow}}n_{i_{\downarrow}}+U%
\underset{j\in B}{\sum}n_{j_{\uparrow}}n_{j_{\downarrow}}+J\underset{\QATOP{%
i\in A}{j\in B}}{\sum}\mathbf{S}_{i}\cdot\mathbf{S}_{j}-\mu N_{e}.
\end{equation}
where the labels $A$ and $B$ represent two interpenetrable sublattices, each
containing half $(N/2)$ of available atomic sites. The first term represents
single-particle hopping of electrons between the sublattices (nearest
neighbors), the second and the third express the intraatomic interaction of
the same magnitude on \textit{all} sites, the fourth includes the Heisenberg
exchange between the sublattices, and $(-\mu N_{e})$ is the reference energy
with $\mu$ being the chemical potential, and $N_{e}(\leq N)\,$the total
number of fermions. The Heisenberg term has been added only to provide an
illustration of the concept of the exchange field coming from the electron
correlations (they will add to one another).

In the mean-field (saddle-point) approximation for the slave bosons, the
rotationally invariant approach of Li et al. \cite{6} and the original
Kotliar-Ruckenstein \cite{1} formulations can be brought to an equivalent
form \cite{7}. Explicitly, the six bosons $e,p_{\binom{A}{B}\uparrow },p_{%
\binom{A}{B}\downarrow }$ and $d$ appearing in the approach fulfill the
following five constrains in the mean-field approximation 
\begin{equation}
e^{2}+p_{\binom{A}{B}\uparrow }^{2}+p_{\binom{A}{B}\downarrow }^{2}+d^{2}=1,
\label{con1}
\end{equation}
\begin{equation}
f_{\binom{A}{B}\mathbf{k}\sigma }^{\dagger }f_{\binom{A}{B}\mathbf{k}\sigma
}=p_{\binom{A}{B}\sigma }^{2}+d^{2},  \label{con2}
\end{equation}
where the subscript ${\binom{A}{B}}$ characterizes the sublattices, and
fermion operators $f_{\mathbf{k}\sigma }\, $and $f_{\mathbf{k}\sigma
}^{\dagger }$ represent the new (quasiparticle) fermion operators appearing
in the theory. The effective Hamiltonian in the saddle-point approximation
takes the form in reciprocal ($\mathbf{k}$)\textbf{\ }space 
\begin{align}
H& =\sum_{\mathbf{k}\sigma }q\varepsilon _{\mathbf{k}}\left( f_{A\mathbf{k}%
\sigma }^{\dagger }f_{B\mathbf{k}\sigma }+f_{B\mathbf{k}\sigma }^{\dagger
}f_{A\mathbf{k}\sigma }\right) +  \label{Hf} \\
& \sum_{\mathbf{k}\sigma }f_{A\mathbf{k}\sigma }^{\dagger }f_{A\mathbf{k}%
\sigma }\left( \lambda _{\sigma }^{(2A)}-\mu -\frac{1}{2}\sigma
Jzm_{B}\right) +  \notag \\
& \sum_{\mathbf{k}\sigma }f_{B\mathbf{k}\sigma }^{\dagger }f_{B\mathbf{k}%
\sigma }\left( \lambda _{\sigma }^{(2B)}-\mu +\frac{1}{2}\sigma
Jzm_{A}\right) +  \notag \\
& NUd^{2}+\frac{N}{2}Jzm_{A}m_{B}+H_{con},  \notag
\end{align}
where $H_{con}$ contains the constraints appearing in the theory in the form
of Lagrange multipliers 
\begin{align}
H_{con}& =\frac{N}{2}\lambda ^{(1A)}\left( e^{2}+d^{2}+p_{A\uparrow
}^{2}+p_{A\downarrow }^{2}-1\right) -  \label{Hcon} \\
& \frac{N}{2}\lambda _{\uparrow }^{(2A)}\left[ \left( p_{A\uparrow
}^{2}+d^{2}\right) +\lambda _{\downarrow }^{(2A)}\left( p_{A\downarrow
}^{2}+d^{2}\right) \right] +  \notag \\
& \frac{N}{2}\lambda ^{(1B)}\left( e^{2}+d^{2}+p_{B\uparrow
}^{2}+p_{B\downarrow }^{2}-1\right) -  \notag \\
& \frac{N}{2}\left[ \lambda _{\uparrow }^{(2B)}\left( p_{B\uparrow
}^{2}+d^{2}\right) +\lambda _{\downarrow }^{(2B)}\left( p_{B\downarrow
}^{2}+d^{2}\right) \right]  \notag
\end{align}
The summation over ($\mathbf{k}$\textbf{)} comprises the states in the
halved Brillouin zone and $z$ is the number of nearest neighbors. The
Lagrange multipliers $\lambda ^{(1A)}$ and $\lambda ^{(1B)}$ correspond to
the constraint (\ref{con1}) and, since they are spin symmetric we can put $%
\lambda ^{(1A)}=\lambda ^{(1B)}=\lambda ^{(1)}.$ The spin-dependent Lagrange
multipliers $\lambda _{\sigma }^{(2A)}\,$ and $\lambda _{\sigma }^{(2B)}$
represent spin-resolved constraint (\ref{con2}). In the antiferromagnetic
(AF) case the magnetic moments $m_{B}\equiv \langle p_{iB\uparrow
}^{2}-p_{iB\downarrow }^{2}\rangle $ $\,$ have the opposite sign, i.e. $%
m_{A}=-m_{B}\equiv m.\,$ This means that the constraints $\lambda _{\sigma
}^{(2)}$ obey the relations $\lambda _{\sigma }^{(2A)}=\lambda {-\sigma }%
^{(2B)}\equiv \lambda _{\sigma }^{(2)}.$ The narrowing factor $q$
renormalizing the bare band energy $\varepsilon _{\mathbf{k}}$ assumes the
form 
\begin{equation}
q=\frac{ep_{A\sigma }+dp_{A\overline{\sigma }}}{\sqrt{\left(
1-d^{2}-p_{A\sigma }^{2}\right) \left( 1-e^{2}-p_{A\overline{\sigma }%
}^{2}\right) }}\frac{ep_{B\sigma }+dp_{B\overline{\sigma }}}{\sqrt{\left(
1-d^{2}-p_{B\sigma }^{2}\right) \left( 1-e^{2}-p_{B\overline{\sigma }%
}^{2}\right) }}.  \label{q1}
\end{equation}

The Hamiltonian (\ref{Hf}) can be diagonalized with the help of the
Bogolyubov transformation \cite{1}. In effect, the free energy functional of
the system (per atom) takes the form 
\begin{align}
\frac{F}{N} & =-\frac{k_{B}T}{N}\underset{\mathbf{k}\sigma}{\sum}\ln\left[
1+e^{-\beta\left( E_{\mathbf{k}}-\mu_{eff}\right) }\right] -\frac{k_{B}T}{N}%
\underset{\mathbf{k}\sigma}{\sum}\ln\left[ 1+e^{-\beta\left( -E_{\mathbf{k}%
}-\mu_{eff}\right) }\right] \quad  \label{Hb} \\
& +\frac{1}{2}Jzm^{2}+Ud^{2}+\mu\frac{N_{e}}{N}+H_{con},  \notag
\end{align}
where now 
\begin{equation}
H_{con}=\lambda^{(1)}\left(
e^{2}+p_{\uparrow}^{2}+p_{\downarrow}^{2}+d^{2}-1\right) -\lambda\left(
p_{\uparrow}^{2}+p_{\downarrow}^{2}+2d^{2}\right) +\Delta m.
\end{equation}
Here the quasiparticle energies are $E_{\mathbf{k}}=\sqrt{\left(
q\varepsilon_{\mathbf{k}}\right) ^{2}+\Delta^{2}}$, with the magnetic gap $%
2\Delta=Jzm+\left(
\lambda_{A\uparrow}^{(2)}-\lambda_{A\downarrow}^{(2)}\right)
\equiv\lambda_{\uparrow}^{(2)}-\lambda_{\downarrow}^{(2)} $, and the
effective chemical potential $\mu_{eff}=\mu-\lambda$ with $\lambda=\frac{1}{2%
}\left( \lambda_{\sigma}^{(2)}-\lambda_{\overline{\sigma}}^{(2)}\right) $.
The quantity $\beta_{m}=\frac{1}{2}\left( \lambda
_{\uparrow}^{(2)}-\lambda_{\downarrow}^{(2)}\right) =\frac{1}{2}\Delta$
plays a role of the correlation induced molecular field, since it adds to
the effective Heisenberg field $Jzm/2.$ In the case $J=0$ (taken in the
numerical analysis) $\beta_{m}$ constitutes the entire gap induced by the
magnetic ordering (it is the \textit{magnetic }gap). On the whole, the first
two terms in (\ref{Hb}) provide the contribution to the thermodynamics
coming from the single particle excitations in the magnetic (Slater)
subbands caused by AF ordering and having energies $\pm E_{\mathbf{k}}.$
These quasiparticle energies comprise the effective band narrowing or the
mass renormalization $m^{\ast}/m_{0}=1/q,$ and the molecular field $%
\beta_{m},$ both to be determined in a self-consistent manner detailed
below. The field $\beta_{m}$ arises from the local constraint (\ref{con2}).
Thus, one can say that the molecular field is induced by the correlations.

The functional (\ref{Hb}) must be minimized with respect to all appearing
Bose fields and the chemical potential $\mu.$ Effectively, one can reduce (%
\ref{Hb}) to the form with two variables $x\equiv2\Delta/(Wq)$ and $d$. For
the purpose of simplicity and clarity of our points we take constant density
of states $(\rho(\varepsilon)=1/W$ for $-W/2\leq\varepsilon\leq W/2),\,$ for
which the ground state energy takes the form 
\begin{align}
\frac{E_{G}}{N} & =-\frac{W}{4}q\sqrt{1+x^{2}}+\left( \frac{1-n}{4}\right) q%
\sqrt{\left( 1-n\right) ^{2}+x^{2}}  \label{Eg} \\
& +Ud^{2}+\frac{1}{2}m\left( \frac{1}{2}xq-jm\right) W,  \notag
\end{align}
where $E_{G}$ is the ground-state energy, and $j\equiv Jz/W.$ In this
expression we have already connected $m$ to $x\,$via the relation 
\begin{equation}
m=x\ln\left| \frac{\sqrt{x^{2}+\left( 1-n\right) ^{2}}-\left( 1-n\right) }{%
\sqrt{x^{2}+1}-1}\right| .
\end{equation}
For the sake of completness, we list also the explicit form of the band
narrowing factor for $n=1$: 
\begin{equation}
q=\frac{2\eta}{1-m^{2}}\left[ 1-2\eta+\sqrt{(1-2\eta)^{2}-m^{2}}\right] .
\end{equation}
Note that the variable $x$ has a physical meaning of the ratio of the Slater
gap to the renormalized band energy. In other words, it provides a relative
strength of the effective local field with respect to the renormalized
kinetic energy. The growing ratio $\Delta/(Wq)\,$ drives the system towards
localization induced by the formation of staggered magnetic moments arranged
in two sublattices, whereas the growing ratio $U/W$ drives the system
towards localization independent of magnetic ordering. Therefore, the
present formulation will allow us to single out the contributions coming
from the two factors. The magnetic energy is thus measured with respect to
the band energy $(\sim Wq)$ renormalized by the Coulomb interaction.

\subsection{Asymptotic limits: $U\rightarrow\infty$ and $U\rightarrow0$}

We now show that the results obtained above provide correctly a mean-field
solution in the $U\rightarrow \infty $ limit. For the sake of simplicity
consider the half-filled band case with $J=0$ and for symmetric form of the
density of states, $\rho (\epsilon )=\rho (-\epsilon )$. We can then write
the ground state energy (per site) in the form 
\begin{equation}
\frac{E_{G}}{N}=-2\int_{-W/2}^{0}d\epsilon \;\rho (\epsilon )\sqrt{%
(q\epsilon )^{2}+\Delta ^{2}}+Ud^{2}+\Delta m.  \label{E1}
\end{equation}
For large $U$, the gap $\Delta $ is also large. In that limit the energy has
the form 
\begin{equation}
\frac{E_{G}}{N}=-\frac{q^{2}}{\Delta }<\epsilon ^{2}>+Ud^{2}+\Delta (m-1),
\label{E2}
\end{equation}
where $<\epsilon ^{2}>\equiv \int_{-W/2}^{0}d\epsilon \;\rho (\epsilon
)\epsilon ^{2}$. Minimizing this expression with respect to $\Delta $ we
obtain that 
\begin{equation}
m\simeq 1-\left( \frac{q}{\Delta }\right) ^{2}<\epsilon ^{2}>.  \label{m1}
\end{equation}
Minimization with respect to $m$ and $d^{2}$ gives the relations 
\begin{equation}
\frac{2q_{m}q}{\Delta ^{2}}<\epsilon ^{2}>=1,  \label{m2}
\end{equation}
and 
\begin{equation}
\frac{2q_{d^{2}}q}{U\Delta }<\epsilon ^{2}>=1,  \label{m3}
\end{equation}
where $q_{x}=dq/dx$. Combining Eqs.(\ref{m1})-(\ref{m3}) we obtain that $%
q=2(1-m)q_{m}$, and $\Delta =(q_{m}/q_{d^{2}})U$. Explicitly, from the fact
that $q\simeq 2\eta /(1-m)$ and that numerically $q\approx 1$, we obtain
that $d^{2}\simeq (1-m)/2$, and $\Delta \approx U/2$. Thus finally, for the
featureless form of the density of states we have 
\begin{align}
m& =1-4\frac{<\epsilon ^{2}>}{U^{2}}=\frac{1}{6}\left( \frac{W}{U}\right)
^{2}, \\
d^{2}& =2\frac{<\epsilon ^{2}>}{U^{2}}=\frac{1}{2}\left( \frac{W}{U}\right)
^{2},
\end{align}
and 
\begin{equation}
\frac{E_{G}}{N}=-2\frac{<\epsilon ^{2}>}{U}.
\end{equation}
In other words, in the $U\rightarrow \infty $ limit the Hubbard model
reduces to the Heisenberg model \cite{2} with the Hubbard gap $\Delta =U$,
since in the mean-field approximation the ground-state energy is then given
by the kinetic exchange contribution \cite{2} 
\begin{align}
\frac{E_{G}}{N}& =-\frac{4}{UN}\sum_{i_{A}j_{B}}t_{i_{A}j_{B}}t_{j_{B}i_{A}}%
\left\langle \mathbf{S}_{i}\cdot \mathbf{S}_{j}-\frac{1}{4}\right\rangle 
\notag \\
& \approx -\frac{2}{UN}\sum_{i_{A}j_{B}\sigma }t_{i_{A}j_{B}}t_{j_{B}i_{A}}=-%
\frac{1}{U(N/2)}\sum_{\mathbf{k}}\epsilon _{\mathbf{k}}^{2}  \notag \\
& =-\frac{1}{U}\int_{-W/2}^{0}d\epsilon \;\epsilon ^{2}\rho (\epsilon
)\equiv -2\frac{\left\langle \epsilon ^{2}\right\rangle }{U},
\end{align}
with $\rho (\epsilon )=\frac{1}{N/2}\sum_{\mathbf{k}}\delta (\epsilon
-\epsilon _{\mathbf{k}})$. Also, the magnetic gap reduces to the atomic
value of the Hubbard gap, as $2\Delta \rightarrow U$ with $<\epsilon
^{2}>/U\rightarrow 0$.

The Hartree-Fock $(U\rightarrow0)$ limit is recovered once one notices that
the method has been constructed to obtain $q=1$ in the weak coupling limit 
\cite{6}, \cite{7}. Under that circumstance Eq.(\ref{Eg}) reduce to the
usual Hartree-Fock form if we assume that now $x=2\Delta/W\ll1$. This limit
was checked out also numerically, but the results are not reproduced here.

\section{Numerical analysis}

In Fig.1 we have displayed both the effective magnetic (Slater-type) gap $%
2\Delta$ and the Mott-Hubbard gap for the paramagnetic phase, both for $n=1$
\begin{figure}
[ptb]
\begin{center}
\fbox{\includegraphics[
height=2.1828in,
width=3.1661in
]%
{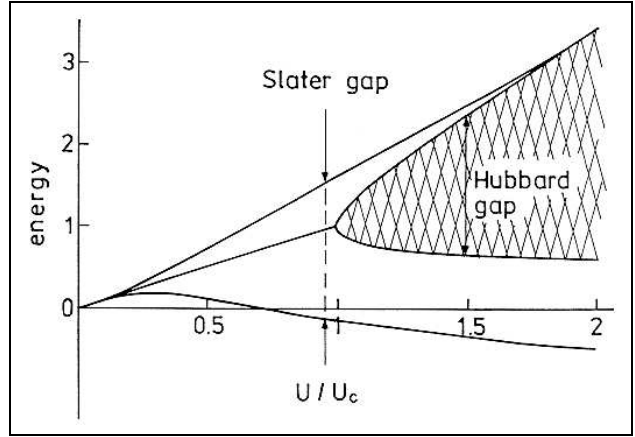}%
}\caption{The Slater gap $2\Delta/W$ and the Hubbard gap as the functions of
interaction strength. The Slater gap merges with the Mott-Hubbard gap as
$U/U_{c}\rightarrow\infty.$}%
\end{center}
\end{figure}
Those characteristics are plotted for the ground state. The chemical
potential is then $\mu(T=0)=\lambda.$ The Mott- Hubbard gap is expressed
through the difference in the chemical potential in the paramagnetic case $%
(\Delta=0)$ for $n=1.001\,$(the upper part) and for $n=0.999$ (the lower
part) and was discussed earlier \cite{8}. For $n=1$ the Slater split-band
picture appears for arbitrary small $U$ and $\Delta$ increases with
increasing $U/U_{c}$, where $U_{c}=8|\bar{\epsilon}|$. In the limit $%
U/U_{c}\sim1$ the gap is composed of the Slater and the Mott-Hubbard parts,
and when $\rightarrow \infty$ the former merges gradually with the latter.
This can be seen explicitly in Fig.2,
\begin{figure}
[ptbptb]
\begin{center}
\includegraphics[
height=2.143in,
width=3.384in
]%
{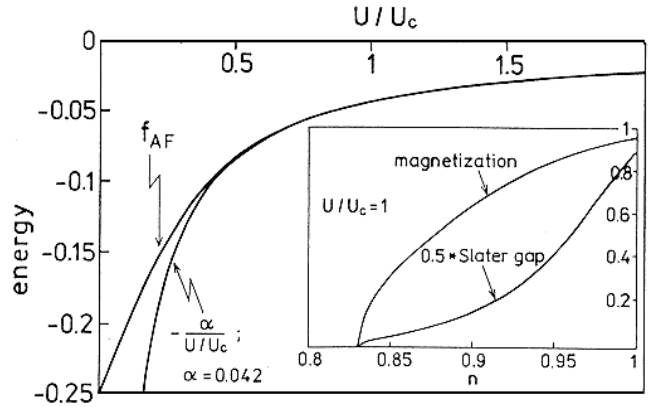}%
\caption{Ground state energy for AF state and $n=1 $ versus $U/U_{c}.$ The
lower curve is the fit to the expression $(-0.042\times U_{c}/U).\,$The inset
displays the difference in behavior of magnetic moment $m$ and half of the
Slater gap $(\Delta/W),$ both plotted as a function of the band filling.}%
\end{center}
\end{figure}
where we have shown the ground state
energy $E/(WN)\,$versus $U/U_{c}.$ In the strong-correlation limit the
energy is determined by the kinetic-exchange contribution $\sim1/U$ \cite{2}%
, as shown explicitly. The fitted coefficient $\alpha=0.042$ to the
numerical results cannot be attributed to any particular 3d structure, since
we have used in numerical calculation a constant value of $\rho(\epsilon)$.
The energies of para- and antiferro-magnetic states merge for $n=1$ and in
the $U\rightarrow \infty$ limit. The inset illustrates another interesting
characteristic of the solution, namely the magnetic gap \textit{is not }%
proportional to the magnetization, as one would expect from the Hartree-Fock
(Slater) solution. In other words, the molecular field $\beta_{m}$ is a
nonlinear function of $m,$ since from the condition $\partial
f/\partial\Delta=0$ we obtain the relation $2\Delta/W=mq/\sqrt{1-m^{2}}.$ $%
\, $ This is the reason why the antiferromagnetic gap for almost localized
fermions cannot be regarded as just the Slater gap. Also, the AF solution
disappears at the critical band filling $n\equiv n_{c}\simeq0.83.$

To visualize the difference between the magnetic gap and the magnetization
we have plotted in Fig.3 both quantities as a function of the interaction
strength $U/U_{c}\equiv U/2W$ $\,$for different band filling parameter $n.$
\begin{figure}
[ptb]
\begin{center}
\includegraphics[
height=3.1099in,
width=2.6152in
]%
{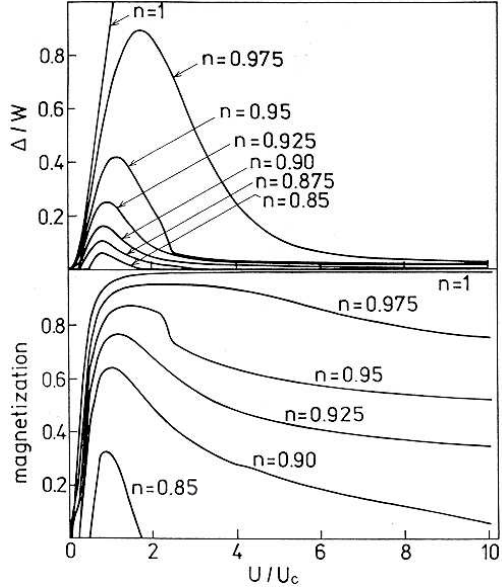}%
\caption{The Slater gap parameter $\Delta/W\,$(top panel) and the magnetic
moment $m=\langle n_{i\uparrow}-n_{i\downarrow}\rangle${\ }(bottom) versus
$U/U_{c}$ and for different values of $n.$\noindent}%
\end{center}
\end{figure}
While for $n=1$ the magnetic moment saturates gradually with growing $%
U/U_{c},$ $\Delta$ keeps increasing. The magnetic gap for $n=1$ increases
and eventually $\Delta\sim U$; this circumstance indicates again that the
magnetic gap merges with the Hubbard gap, which can be estimated
analytically and is $\approx$ $-W+4UW^{2}/(z(U^{2}+W^{2}))\rightarrow
U(1-W/U)$ for $W/U\rightarrow0$.

The double occupancy probability $d^{2}=\langle n_{i\uparrow}n_{i\downarrow
}\rangle$ is shown in Fig.4 for different band fillings.
\begin{figure}
[ptb]
\begin{center}
\includegraphics[
height=2.3203in,
width=3.4878in
]%
{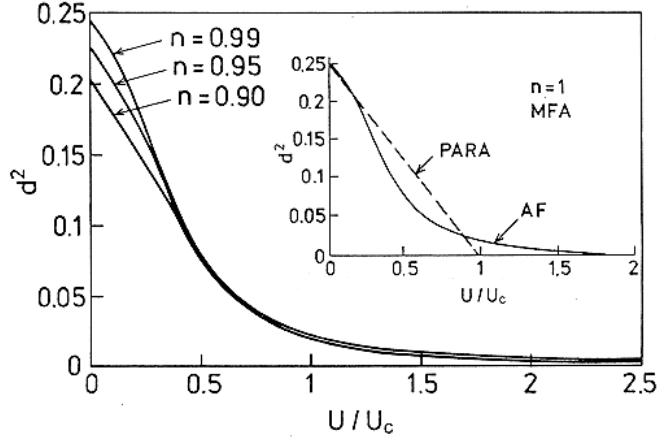}%
\caption{The double occupancy $d^{2}=\langle n_{i\uparrow}n_{i\downarrow
}\rangle$ vs $U/U_{c},\,$and for the different $n$ values.The inset display
the difference in behavior for para- and antiferro-magnetic cases for $n=1.$}%
\end{center}
\end{figure}
It decreases
continuously with growing $U/U_{c}$, i.e. the charge fluctuations are
gra\-dual\-ly suppressed, while the magnetic moment grows (cf. Fig.3). The
difference in the behavior of $d^{2}\,$and $m\,$is caused by the
circumstance that the $d^{2}$ is of intraatomic nature, whereas $m$ is
determined from the competition between the magnetic energy $\sim\beta{m}%
^{2} $ (also of intraatomic nature) and the renormalized band energy $\sim
Wq\,$. The inset to Fig.4 exemplifies the difference between the diminution
of $d^{2}$ with growing $U/U_{c}$ for $n=1$ in two situations: For the
paramagnetic (PARA) case $d^{2}\equiv0$ for $U\geq U_{c};\,$ this feature is
concurrent with the well-known effective mass divergence at the Mott-Hubbard
localization boundary (the \textit{Brinkman-Rice point}). This divergence 
\textit{does not emerge} in the antiferromagnetic state as $d^{2}$
approaches zero gradually, in the same manner, as $m$ approaches saturation $%
m\rightarrow1).$

In Fig.5 we have displayed the inverse band narrowing factor $q^{-1}$ (for $%
n=1$) which turns into the effective mass renormalization $%
m^{\ast}/m_{0}=1/q $ (for $n<1$) as a function of $/U_{c}\,$ and for
different $n\,\ $\ values.
\begin{figure}
[ptb]
\begin{center}
\includegraphics[
height=2.2969in,
width=3.3831in
]%
{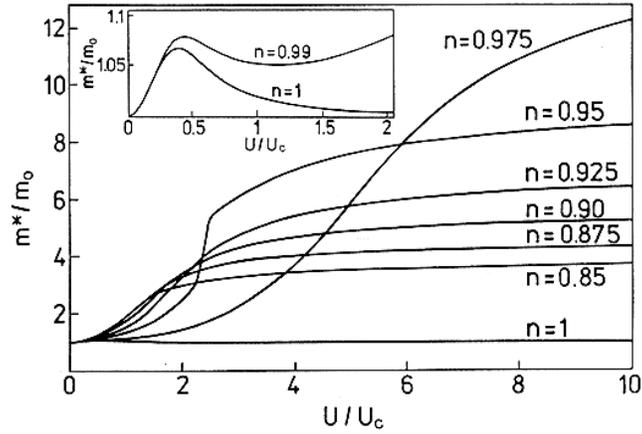}%
\caption{The effective mass enhancement $m^{\ast}/m_{0}$ (with respectto the
band value $m_{0}$) vs $U/U_{c}$ and for the $n$ values shown. The inset shows
a rather weak enhancement close to the Mott-Hubbard limit.}%
\end{center}
\end{figure}
Again, the inset illustrates the difference with
the $n=1$ case. One should note that the enhancement factor in AF state is
very small compared to that in PARA state, which is equal to $1/q=\left[
1-\left( U/U_{c}\right) ^{2}\right] ^{-1}.\,\ $The difference between AF and
PARA states diminishes with decreasing $n,\,$as in that situation the
magnetic moment is reduced rapidly. So, the weak band narrowing in the $n=1$
case can be associated with the circumstance that the Fermi level falls in
the gap. This is the reason why $m^{\ast}/m_{0}$ raises rather steeply
around $n\simeq0.95$. Thus, the appearance of the itinerant
antiferromagnetism with an isotropic gap changes the Brinkman-Rice scenario
for the Mott-Hubbard transition, as it has been underlined before \cite{9}.
Also, the physical parameters $d^{2},1-m$ and $1-n$ are all of the same
magnitude. This is easy to envisage when estimating e.g. the band narrowing $%
q,$ which is in the AF state roughly $sim2d^{2}/(1-m^{2})$ and is of the
order of unity.

In Fig.6 we have displayed the stability regime $(n,U/U_{c})$ of the AF
phase.
\begin{figure}
[ptb]
\begin{center}
\includegraphics[
height=1.9934in,
width=3.3883in
]%
{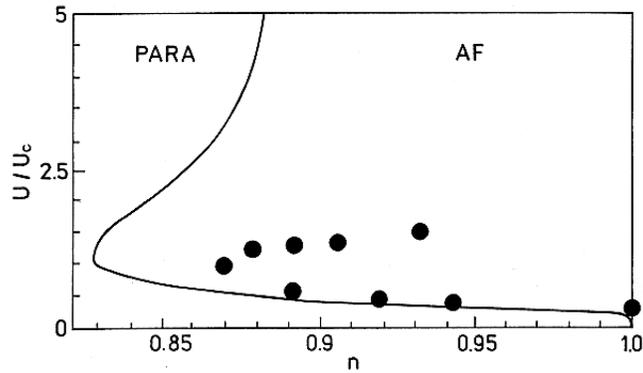}%
\caption{The stability regime of AF solution; the inset: results of
Monte-Carlo calculations in the $d\rightarrow$limit [8]\noindent\ }%
\end{center}
\end{figure}
The full circles has been obtained \cite{10} in the limit of infinite
dimension with the help of quantum Monte Carlo simulation. One should note
the range of the filling $n$ of stable AF phase is the broadest for $%
U/U_{c}\sim1,$ i.e. when the molecular field is the strongest (cf. Fig.3).
Note also that the Monte-Carlo results did note provide the asymptotic
behavior for $n\rightarrow1$, as it does not reduce correctly to the
Hartree-Fock limit.

For the sake of completeness we display in Fig.7 the ground state energy as
a function of $U/U_{c}$, for different $n\,$close to the half filling.
\begin{figure}
[ptb]
\begin{center}
\includegraphics[
height=1.9545in,
width=2.943in
]%
{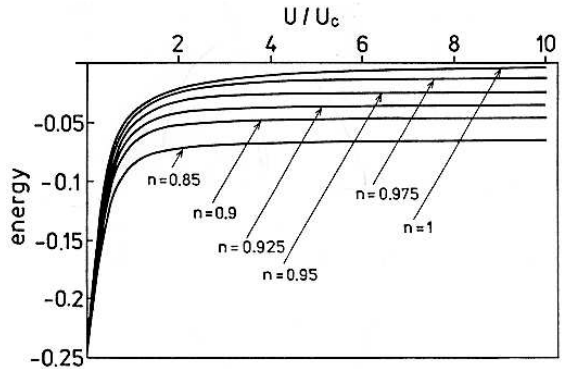}%
\caption{Ground state energy vs $U/U_{c}\, $for different band filling $n$. In
each case the energy is $\sim U $ for $U\lesssim0.5U_{c},$ and $\sim1/U$ for
$U>U_{c}.$ }%
\end{center}
\end{figure}
In
each case (for $n<1$) the system contains the contribution $\sim U$ for $%
/U_{c}\lesssim0.5\,\,$and the contribution $\sim1/U$ for $U/U_{c}>1.$ In
other words, the solution interpolates between the Hartree-Fock and the
kinetic exchange limits. Moreover, the shift of the curves with diminishing $%
\,$ $n$ $\,$ in the $U\rightarrow\infty$ limit is due to the presence of the
band term proportional to $n(1-n).$ One should note that in the paramagnetic
state the direct Coulomb interaction contribution $(Ud^{2})$ competes with
band energy $(-Wq/4).$ In the antiferromagnetic state the local magnetic
contribution $(-|\beta_{m}|m)$ is of the opposite sign than the Coulomb
part. However, it changes also the band energy. To specify the role of the
staggered field we have plotted in Fig.8 the difference $(Ud^{2}-|%
\beta_{m}|m)/W$ as a function of $U/U_{c}.$
\begin{figure}
[ptbptb]
\begin{center}
\includegraphics[
height=1.8412in,
width=3.1998in
]%
{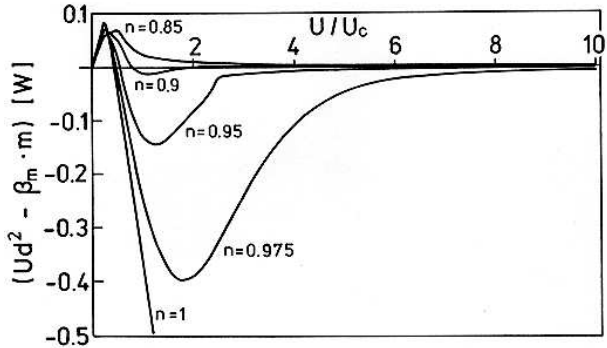}%
\caption{Comparison of the magnetic $(-\beta_{m}m)\;$and the Coulomb
$(Ud^{2})\;$terms in the ground state energy versus $U/U_{c}.$ }%
\end{center}
\end{figure}
The magnetic contribution
exceeds the Coulomb part for $U>0.5U_{c}$ (i.e. away from the Hartree-Fock
limit), and is particularly strong (and comparable to the band energy) in
the regime $U\sim U_{c},$ where the Mott-Hubbard localization would take
place for the paramagnetic case. This circumstance indicates why the
metal-insulator transition present at $T=0$ in the paramagnetic phase is
wiped out when the itinerant AF sets in. Simply, the magnetic contribution $%
(\sim\beta_{m}m)$ freezes gradually the spatially separated electrons in the
antiferromagnetic phase, taking also advantage of nonvanishing kinetic
energy (the kinetic exchange, $\sim<\epsilon^{2}/U$, contribution!). In the
PARA phase we have instead a direct competition between band and Coulomb
energies.

To summarize, the Mott-localization is achieved gradually at $T=0$ in the AF
state. In other words, the present approach provides a crossover behavior
from Slater to Mott insulator, as shown e.g. in Fig.2. The same holds true
even when we include the intersite exchange $(J>0),$ which may originate
from the quantum Gaussian fluctuations. The continuous evolution with
growing $U/W\,$does not preclude the first order transition at nonzero
temperature, as has been demonstrated some time ago for the paramagnetic
state \cite{11} and subsequently reconfirmed in the limit $%
d\rightarrow\infty $ \cite{12}. The correlated state builts in gradually
with increasing $U/U_{c}$, as can be seen from the gradual increase of the
mass enhancement in $n<1\,$case, i.e. when the ground state is always
metallic. The same holds true for the ground state energy shown in Fig.7.

\section{Discussion and conclusions}

We have addressed the question of crossover from Slater (Hartree-Fock) to
Mott-Hubbard (atomic) picture in the half-filled band case, as well as have
analyzed the behavior of quasiparticle physical properties in AF state in
the half- and non half-filled-band cases. Although our analysis is based on
the saddle-point solution within the slave-boson functional-integral
approach, the results can serve as a mean-field analysis, since they
interpolate between those in the Hartree-Fock approximation in the limit $%
U\rightarrow 0$ and those in mean-field approximation for the Heisenberg
model (for $n=1)$ in the $U\rightarrow \infty $ limit. They also represent
basis for inclusion of Gaussian fluctuations \cite{13} in a magnetically
ordered (AF)\ state close to the Mott-Hubbard localization. However, one
should realize that for the half-filled band case the effective mass is only
weakly renormalized so the renormalization factor $z_{k_{F}}=(m_{0}/m^{\ast
})\sim 1$. Also, for $n\neq 1,$ $z_{k_{F}}$ remains always finite so,
perhaps the role of the quantum fluctuations is not as crucial for AF state,
as it is for the paramagnetic state. In any case, it will be much more
involved that in the paramagnetic case \cite{13}. The full analysis of the
Mott-Hubbard boundary should also include the disordered local-moment phase 
\cite{14}, so far not included within the present scheme.

The physical meaning of the results obtained within the slave-boson approach
(SBA) can be characterized as follows. It is well known \cite{1}, \cite{6} - 
\cite{8}, \cite{13} that the saddle point solution of the approach
reproduces the results of the Gutzwiller approach (GA), as far as the
overall (ground-state) behavior is concerned. Moreover, SBA leads to an
improvement of the GA by incorporating both the quasiparticle picture of
those systems and the quantum Gaussian fluctuations (not considered in this
paper). In general, the pseudo-fermion fields $\{f_{i\sigma }\}$
representing the quasiparticle states are in one-to-one correspondence to
the original fermion fields in the physical Fock space. The mapping is quite
obvious within the above analysis. Specifically, the quasiparticle energy $%
E_{\mathbf{k}}=\sqrt{\left( q\epsilon _{\mathbf{k}}\right) ^{2}+\Delta ^{2}}%
\,$ leads to the following density of states 
\begin{equation}
\rho (E)=\left. \frac{2\left| E\right| }{\sqrt{E^{2}-\Delta ^{2}}}\rho
^{0}(\epsilon )\right| _{\epsilon =\sqrt{E^{2}-\Delta ^{2}}},  \label{ro}
\end{equation}
where $\rho ^{(0)}(\epsilon )\,$is the bare density of states in PM phase.
Thus the enhancement factor due to the correlations is distinct (and
disappear in the first factor) from the change of the density of states
caused by the appearance of the magnetic gap. By analogy, for the
paramagnetic state ($\Delta \rightarrow 0$), for which $E_{\mathbf{k}}$ can
be written as 
\begin{equation}
E_{\mathbf{k}}=q\epsilon _{\mathbf{k}}\equiv \epsilon _{\mathbf{k}}+\Sigma
(\epsilon _{\mathbf{k}}),
\end{equation}
and where the self-energy $\Sigma (\omega )=(1-q)\omega $ leads to the mass
enhancement $m^{\ast }/m=(1-\partial \Sigma /\partial \omega )^{-1}=1/q,$ we
cannot single out the factor ($1/q$) in (\ref{ro}) as the corresponding
enhancement also in AF state. In the doubly-degenerate-band case (for $n=1$)
and under the same-type of approximation scheme, the role of the magnetic
gap is played by the gap formed by an alternant orbital ordering in the
ferromagnetic state \cite{3}. The gap $(\delta )$ in the latter case is
reproduced in Fig.9 for the magnitude of the intraatomic (Hund's-rule)
exchange $J/U=0.2$.
\begin{figure}
[ptb]
\begin{center}
\includegraphics[
height=2.4613in,
width=2.8513in
]%
{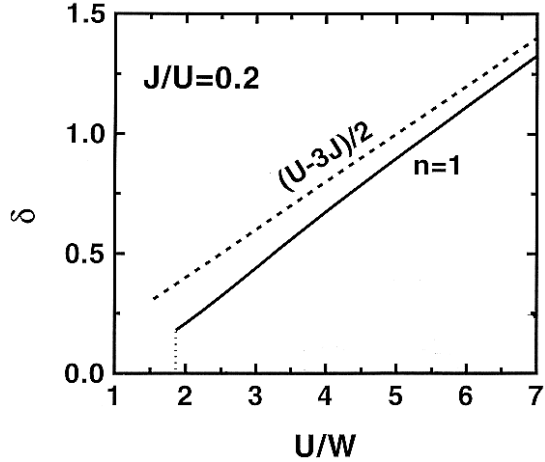}%
\caption{The half-gap parameter $\delta$ (in units of the bare bandwidth $W$)
vs $U/W$ for a quarter-filled doubly-degenerate band. The Hund's rule exchange
is $J=0.2U$.}%
\end{center}
\end{figure}
Again, the gap approaches asymptotically the
Mott-Hubbard gap, which in this case \cite{3} is equal to $(U-3J)/2$.
However, here there is a critical value of $U/W$, at which the system
exhibits antiferromagnetic orbital ordering. In the half-filled case for the
doubly degenerate case $(n=2)$ the gap appears for an arbitrary small $U$ 
\cite{15} in the AF case and for the critical value of $U$ for the
paramagnetic phase \cite{15}.

One should also note that within the present SB\ scheme the
antiferromagnetic Slate-type state (AFS) evolves for $n=1$ gradually into
antiferromagnetic Mott (AFI)\ state with increasing $U/W$ ratio. For $n<1$,
we observe only antiferromagnetic metallic (AFM) state. This is in contrast
with the results obtained with the composite-operator method \cite{16}, as
well as with an original analysis \cite{17} where a phase border line
between itinerant- and localized-staggered moment-bearing phases is drawn.
Localized-moment phase for $n=1-\delta ,$ $\delta <<1$ can be obtained only
when the polaronic effects due to antiferromagnetic moments surrounding the
hole in the Mott insulator are included \cite{18}. Obviously, such ASF-AFI
border line at temperature $T>0$ (for $n=1$)\ is present \cite{11} and is
induced by the difference in entropy of AFS and AFI states.

The computational details of the present work are available on request \cite
{19}.

\bigskip 

\textbf{Acknowledgment} J.S. is grateful to Claudine Lacroix and M. Avignon
for discussions during his stay at DRFMC-CENG in Grenoble. This research was
supported by KBN Grant No. 2 P03B 050 23. Two of the authors (ML and JS)
acknowledge also the support of the Project franco-polonais ''Polonium'' in
the period 1998-2000.

{\large \textbf{Figure Captions}}

\begin{itemize}
\item  Fig.1. The Slater gap $2\Delta /W$ and the Hubbard gap as the
functions of interaction strength. The Slater gap merges with the
Mott-Hubbard gap as $U/U_{c}\rightarrow \infty .$

\item  Fig.2. Ground state energy for AF state and $n=1$ versus $/U_{c}.$
The lower curve is the fit to the expression $(-0.042\times U_{c}/U).\,$The
inset displays the difference in behavior of magnetic moment $m$ and half of
the Slater gap $(\Delta /W),$ both plotted as a function of the band filling.

\item  Fig.3. The Slater gap parameter $\Delta /W\,$(top panel) and the
magnetic moment $m=\langle n_{i\uparrow }-n_{i\downarrow }\rangle ${\ }%
(bottom) versus $U/U_{c}$ and for different values of $n.$\noindent 

\item  Fig.4. The double occupancy $d^{2}=\langle n_{i\uparrow
}n_{i\downarrow }\rangle $ vs $U/U_{c},\,$and for the different $n$
values.The inset display the difference in behavior for para- and
antiferro-magnetic cases for $n=1.$

\item  Fig.5. The effective mass enhancement $m^{\ast }/m_{0}$ (with
respectto the band value $m_{0}$) vs $U/U_{c}$ and for the $n$ values shown.
The inset shows a rather weak enhancement close to the Mott-Hubbard limit.

\item  Fig.6. The stability regime of AF solution; the inset: results
ofMonte-Carlo calculations in the$d\rightarrow $limit \cite{8}.\noindent 

\item  Fig.7. Ground state energy vs $U/U_{c}\,$for different band filling $n
$. In each case the energy is $\sim U$ for $U\lesssim 0.5U_{c},$ and $\sim
1/U$ for $U>U_{c}.$

\item  Fig.8. Comparison of the magnetic $(-\beta _{m}m)\;$and the Coulomb $%
(Ud^{2})\;$terms in the ground state energy versus $U/U_{c}.$

\item  Fig.9. The half-gap parameter $\delta $ (in units of the bare
bandwidth $W$) vs $U/W$ for a quarter-filled doubly-degenerate band. The
Hund's rule exchange is $J=0.2U$.
\end{itemize}


\begin{thebibliography}{99}
\bibitem{1}  See: J. des Cloizeaux, J. Phys. Radium. \textbf{20}, 606
(1959), and D.R. Penn, Phys. Rev.\textbf{142}, 350 (1966), - for the
stability of antiferromagnetic (AF)\ phases in the Hartree-Fock (HF)
approximation. The stability of AF in the present context (i.e. in the
Gutzwiller approximation) has been obtained in: K.Kubo and M. Uchinami,
Prog. Theor. Phys. \textbf{54}, 1289 (1975). For a comparative analysis see:
e.g. A.M. Ole\'{s} and J. Spa\l ek Z. Phys.B \textbf{44}, 177 (1981); see
also: G. Seibold et al., Phys. Rev. B \textbf{57}, 6937 (1998). The phase
diagram recovering both HF and strong correlation limits correctly has been
discussed in G. Kotliar and A.E. Ruckenstein, Phys. Rev. Lett. \textbf{57},
1362 (1986)-within the slave-boson approach; and in: W. Metzner and D.
Vollhardt, Phys. Rev. Lett. \textbf{62}, 324 (1989); P. Fazekas, B. Menge
and E. M\"{u}ler-Hartmann, Z. Phys. \textbf{78}, 69 (1990)-in the limit of
infinite dimension $d\rightarrow \infty )$. The corresponding discussion in
two-spatial dimensions was performed in: E. Arrigoni and G.C. Strinati,
Phys. Rev. B\textbf{44}, 7455 (1991); W. Ziegler et al., Phys. Rev. B 
\textbf{53}, 1231 (1996).

\bibitem{2}  In that respect the AF stability should be considered for $\neq
1$ against the onset of the ferromagnetism, for which the Pauli principle
plays the same role as $U\rightarrow \infty $ for antiferromagnetism.
However, in the large-$U$ limit, the virtual hopping processes, particularly
for $n=1$, lead to a stable AF state; see e.g. J.Spa\l ek, \ A.M. Ole\'{s},
and K.A. Chao, phys. stat. solidi (b) \textbf{108}, 329 (1981). Cf. also: K.
A.Chao, J. Spa\UNICODE{0x142}{}ek, and A.M. Ole\'{s}, J. Phys. C\textbf{10},
L271 (1977), where the effective kinetic exchange has been derived in the $%
W/U\ll 1$ limit for an arbitrary $n$. Asymptotic values of the magnetic
moment and the molecular-field parameter $\beta $ are discussed in: B. M\"{o}%
ller et al., J. Phys.: Condens. Matter \textbf{5}, 4847 (1993).

\bibitem{3}  A. Klejnberg and J. Spa\UNICODE{0x142}{}ek, Phys. Rev. B\textbf{%
61}, 15542 (2000); \textit{ibid.}, \textbf{57}, 12041 (1998).

\bibitem{4}  Cf. also in F. Gebhardt, \textit{The Metal-Insulator \
Transition}, Springer Tracts in Modern Physics, vol. \textbf{137} (1997).

\bibitem{5}  See e.g.: P. Korbel, J. Spa\UNICODE{0x142}{}ek, W. W\'{o}jcik,
and M. Acquarone, Phys. Rev. B\textbf{52}, R2213 (1995).

\bibitem{6}  T. Li et al., Phys. Rev. B\textbf{40}, 6817 (1989); R. Fr\'{e}%
sard and P. W\"{o}lfle, Int. J. Mod. Phys. B\textbf{6}, 237 (1992); \textbf{6%
}, 3087(E) (1992); W. Ziegler et al., in Ref[1].

\bibitem{7}  J. Spa\l ek and W. W\'{o}jcik, in \textit{Spectroscopy of the
Mott Insulators and Correlated Metals}, Springer Series in Solid State
Sciences, vol. \textbf{119}, pp.41-65 (1995).

\bibitem{8}  M. Lavagna, Phys. Rev. B\textbf{41}, 142 (1990); Int. J. Mod.
Phys. B\textbf{5}, 885 (1991).

\bibitem{9}  X. Yao, J.M. Honig, T. Hogan, C. Kanewurf, and J. Spa%
\UNICODE{0x142}{}ek, Phys. Rev. B\textbf{54}, 17469 (1996).

\bibitem{10}  M. Jarrell, Phys. Rev. Lett. \textbf{69}, 168 (1992); M.
Jarrell and T. Pruschke, Z.Phys. B\textbf{90}, 187 (1993). For the
comparison the men-field slave-boson and the quantum Monte-Carlo approaches
see: L. Lilly et al., Phys. Rev. Lett. \textbf{65}, 1379 (1990).

\bibitem{11}  J. Spa\l ek et. al., Phys. Rev. Lett. \textbf{48}, 729 (1987);
for review see: J. Spa\l ek, J. Solid State Chem. \textbf{88}, 70 (1990).

\bibitem{12}  For review see: A. Georges et. al., Rev. Mod. Phys. \textbf{68 
}, 13 (1996).

\bibitem{13}  The Gaussian fluctuations in PARA state are treated in Ref. 6,
and in: P. W\"{o}lfle and T. Li, Z. Phys. \textbf{78}, 45 (1990); R.
Raimondi and C. Castellani, Phys. Rev. B\textbf{48}, 11453 (1993). For
recent treatment see: R. Fr\'{e}sard and T. Kopp, Nucl. Phys. B \textbf{594}%
, 769 (2001); A. Tandon et al., Phys. Rev. Lett. \textbf{83}, 2046 (1999).

\bibitem{14}  R.M. Ribeiro-Teixeira and M. Avignon, in \textit{New Trends in
Magnetism, Magnetic Materials, and Their Applications, }edited by J. L.
Moran-L\'{o}pez and J.M. Sanchez (Plenum Press, New York, 1994) p. 373.

\bibitem{15}  H. Hasegawa, J. Phys. Soc. Jpn., \textbf{66}, 3522 (1997);
Phys. Rev. B\textbf{56}, 1196 (1997); A. Klejnberg and J. Spa%
\UNICODE{0x142}{}ek, Phys. Rev. B\textbf{57}, 12041 (1998).

\bibitem{16}  A. Avella, F. Mancini and R. M\"{u}nzer, Phys. Rev. B\textbf{63%
}, 245117 (2001).

\bibitem{17}  M. Cyrot, Phil. Mag. \textbf{25}, 1031 (1972), J. Phys (Paris) 
\textbf{33}, 125 (1972); T. Hermann and W. Nolting, J. Magn. Magn. Mat. 
\textbf{170}, 253 (1997).

\bibitem{18}  These effects are considered mainly in the context of large-U
situation for a two dimentional case, cf. e.g. P. Wr\'{o}bel and R. Eder,
Phys. Rev. B\textbf{58}, 15160 (1998), and references therein; cf. also: M.
Imada et al., Rev. Mod. Phys. \textbf{70}, 1039 (1998).

\bibitem{19}  P. Korbel, Ph.D. thesis (Jagiellonian University - Krak\'{o}w,
unpublished).\newpage 
\end{thebibliography}
\end{document}